\documentclass[apjl]{emulateapj}
\usepackage{graphicx,natbib,apjfonts}

\begin{document}

\date{\today}
\title{THE UBIQUITOUS RADIO CONTINUUM EMISSION FROM THE MOST MASSIVE EARLY-TYPE GALAXIES}
\author{
Michael J. I. Brown\altaffilmark{1},
Buell T. Jannuzi\altaffilmark{2},
David J. E. Floyd\altaffilmark{3},
Jeremy R. Mould\altaffilmark{3,4}}
\altaffiltext{1}{School of Physics, Monash University, Clayton, Victoria 3800, Australia}
\altaffiltext{2}{National Optical Astronomy Observatory, Tucson, AZ 85726-6732, USA}
\altaffiltext{3}{School of Physics, University of Melbourne, Parkville, Victoria 3010, Australia}
\altaffiltext{4}{Centre for Astrophysics and Supercomputing, Swinburne University of Technology, Melbourne, Victoria 3122, Australia}

\begin{abstract}
We have measured the radio continuum emission of 396 early-type galaxies brighter than $K=9$, using 
$1.4~{\rm GHz}$ imagery from the NRAO VLA Sky Survey, Green Bank 300-ft Telescope and 64-m Parkes Radio Telescope. 
For $M_K<-24$ early-type galaxies, the distribution of radio powers at fixed absolute magnitude spans 4 orders of 
magnitude and the median radio power is proportional to $K$-band luminosity to the power $2.78\pm 0.16$. The measured 
flux densities of $M_K<-25.5$ early-type galaxies are greater than zero in all cases. It is thus highly likely
that the most massive galaxies always host an active galactic nucleus or have recently undergone star formation.
\end{abstract}

\maketitle

\section{Introduction}
The most massive early-type galaxies reside within dark matter halos with masses of $10^{13}~{\rm M_\odot}$ or 
more \citep[e.g.,][]{bro08}. One may expect the gas within these dark matter halos to radiatively cool, 
gravitationally collapse and form stars \citep[e.g.,][]{whi78}. In contrast to this expectation, 
stellar population synthesis modeling reveals that the bulk of the stars in very massive galaxies 
formed $\simeq 10~{\rm Gyr}$ ago, with relatively little star formation since \citep[e.g.,][]{tin68,tra00}. 
While some early-type galaxies harbor cold gas, the most massive galaxies are typically bathed in 
plasma with temperatures on the order of millions of Kelvin \citep[e.g.,][]{can87}. What is heating the plasma
and regulating  star formation in very massive galaxies has yet to be robustly identified.

In the recent literature, a popular mechanism for heating the plasma is the injection of energy from 
an active galactic nucleus \citep[AGN feedback; e.g.,][]{tab93,cro06}. While a quasar could heat the 
gas within its host galaxy \citep[e.g.,][]{hop06}, powerful quasars are so rare that they  
probably cannot be responsible for the lack of star formation in nearby massive galaxies. For nearby galaxies, 
it is more plausible that low luminosity AGNs (with a high duty cycle) are heating the plasma. 
Bubbles in the distribution of X-ray emitting gas surrounding nearby radio galaxies correspond to the locations of 
radio lobes \citep[e.g.,][]{fab03a,mcn00}. The dissipation of shocks and sound waves resulting from the 
production of these bubbles and jets may inject sufficient energy into the plasma to offset radiative 
cooling \citep{fab03a}. 

Radio observations provide insights into the plausibility and nature of AGN feedback.
We may expect to observe radio emission from all massive galaxies, resulting from star formation
and/or AGNs. Radio emission is associated with the cavities in the X-ray emitting plasma produced by
AGNs and radio emission results from recent star formation \citep{con92}. However, in isolation,
the presence of radio emission from massive galaxies is not proof of AGN regulation of star formation.
For example, AGNs with low power jets \citep[e.g.,][]{bal09} may have little impact on the surrounding plasma.
If there are massive galaxies without radio emission, this may indicate that brief bursts of AGN activity are sufficient
to truncate star formation \citep[e.g.,][]{hop06}. Alternatively, a mechanism other than AGN feedback may be
heating the plasma surrounding galaxies \citep[e.g.,][]{bir03}.

Previous studies show that $\sim 30\%$ of the most massive galaxies are radio continuum 
sources \citep[e.g.,][]{fab89,sad89,wro91,bes05,sha08}. These studies matched optical and radio source catalogs, 
which limits the sample to radio sources that meet conservative signal-to-noise criteria, so catalogs 
are not swamped by noise. (Even when conservative criteria are applied, the tail of the noise distribution
may produce spurious sources.) Recent studies have generally utilized large redshift 
surveys that exclude the nearest galaxies, and thus miss radio sources fainter than $10^{22}~{\rm W~{Hz}^{-1}}$. 
Consequently, the faint radio emission from nearby early-type galaxies has not been completely characterized
by the prior literature.


In this letter we present a study of the 1.4~GHz radio emission from $K<9$ early-type galaxies. The choice of 1.4~GHz is 
pragmatic, as it allows us to utilize existing NRAO VLA Sky Survey \citep[NVSS;][]{con98} and single-dish imagery. 
We assume the radio emission is the consequence of either recent star formation or an AGN. If this assumption holds, 
our conclusions do not depend on which emission process is dominant in these galaxies (i.e., synchrotron, free-free).
Rather than match our early-type galaxies to radio source catalogs alone, we also measure flux densities from radio images. 
We can thus include significant (albeit noisy) information on the 
radio flux densities of early-type galaxies that would have otherwise been excluded from our study. This allows us 
to characterize the very faint radio emission from the most massive early-type galaxies.

\section{The Sample}

Our parent sample is the 2MASS Extended Source Catalog \citep{jar00}, from which we select 
objects with apparent magnitude $K<9$ \citep[dust corrected;][]{sch98}, declination $\delta>-40^{\circ}$ and galactic latitude 
of $|b|>15^{\circ}$. Of the 1107 objects selected with these criteria, 979 have morphologies 
available from the Third Reference Catalog of Bright Galaxies \citep{rc3} while virtually
all of the remaining objects are Galactic.

Our principal sample is the 400 galaxies that are classified as elliptical or lenticular 
galaxies in the RC3 (with T type classifications of  -1 or less). Many of these galaxies 
have redshifts in the RC3 catalog, while the remainder have redshifts provided by 
\citet{fal99}, \citet{huc99}, \citet{weg03}, \citet{jon09} and the NASA Extragalactic Database.
For 170 galaxies we have redshift independent distances from Extragalactic Distance 
Database \citep{tul09}, while for the remaining galaxies we use a Hubble 
constant of $73~{\rm km~s^{-1}~Mpc^{-1}}$ \citep{spe07}. The luminosity distances
of the galaxies span from $0.7$ to $10^2~{\rm Mpc}$

Our principal radio imaging is the $1.4~{\rm GHz}$ NRAO VLA Sky Survey \citep{con98},
which has an angular resolution of $45^{\prime\prime}$ and an RMS noise of $\simeq 0.45~{\rm mJy}$.
As the NVSS can underestimate the flux densities of powerful and extended radio sources, we also measure 
flux densities using lower resolution ($\sim 12^\prime$) imaging from the 300-ft Green Bank and the 
64-m Parkes radio telescopes \citep[][Calabretta in prep.]{con85,con86,bar01}. 

We employed the following method to measure flux densities for each galaxy. At the 2MASS position
of each galaxy we measured a flux density per beam directly from the NVSS images. For those
galaxies with a flux density per beam greater than $2~{\rm mJy}$, we searched for the nearest 
counterpart  in the NVSS catalog within $2^{\prime}$ and used the NVSS catalog deconvolved flux density. 
To account for powerful extended sources, we measured the flux density per beam using Green Bank 300-ft and Parkes 
single dish data, and utilized this flux density measurement if it was greater than $0.6~{\rm Jy}$.
If the flux density per beam measured with single dish imagery was greater than $5~{\rm Jy}$, we utilized 
the integrated flux density rather than the flux density per beam. 

Our flux density measurements are imperfect, but our principal conclusions remain unchanged 
unless gross systematic errors are present. 
To verify that our data was free of such errors, we visually inspected the radio imagery and compared
our flux densities with those from the literature. Visual inspection also revealed that few galaxies
in our sample are extended \citet{far74} class II radio sources. The flux densities of 4 galaxies 
neighboring bright radio sources could not be reliably measured, and excluding these galaxies reduced 
our final sample to 396 early-type galaxies. The properties of the sample 
galaxies are summarized in Table~\ref{table:summary}.

\section{The Radio Flux Density and Luminosity Distributions}

\begin{figure}[hbt]
\begin{center}
\resizebox{3.25in}{!}{\includegraphics{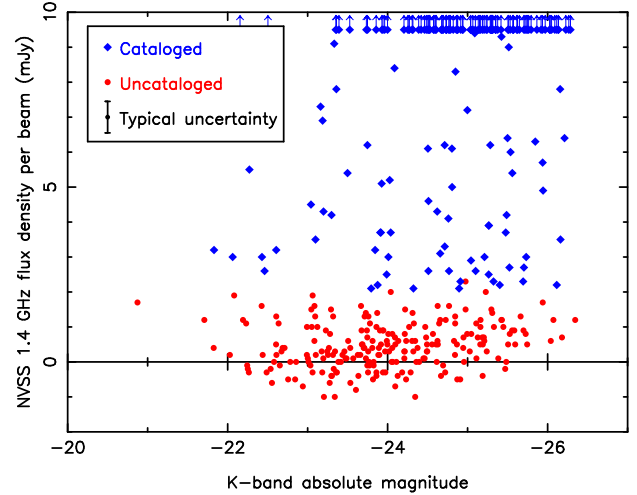}}
\end{center}
\caption{The flux densities of early-type galaxies as a function of $K$-band absolute magnitude.
For the lowest mass galaxies, relatively few galaxies have counterparts in the NVSS catalogs
and most galaxies have flux densities consistent with zero plus a random measurement error.
Galaxies with $M_K<-25.5$ (corresponding to stellar masses of $>2\times 10^{11}~M_\odot$) 
have counterparts in the NVSS catalogs or flux densities greater than zero. As random errors
will broaden the flux density distribution, it is likely that all $M_K<-25.5$ early-type
galaxies in our sample are radio continuum sources with flux densities of $\gtrsim 1~{\rm mJy}$.}
\vspace*{0.5cm}
\label{fig:fluxes}
\end{figure}

The $1.4~{\rm GHz}$ flux densities of the 396 $K<9$ early-type galaxies are 
plotted as a function of $K$-band absolute magnitude in Figure~\ref{fig:fluxes}. 
For the lowest mass galaxies, with $M_K\sim -22.5$, only a small 
fraction have counterparts in the NVSS catalogs and most galaxies have flux densities 
consistent with zero plus a random measurement error.
In contrast, all 59 of the $M_K<-25.5$ early-type galaxies (corresponding to stellar masses 
of $\gtrsim 2.5\times 10^{11}~M_\odot$) have measured flux densities greater than zero 
and just two of the 61 $-25.0<M_K<-25.5$ early-type galaxies have measured flux densities below zero. 
As random  errors will result in measured flux densities scattered above and below
the true flux densities, it is highly likely that all $M_K<-25.5$ early-type galaxies in our sample are 
radio continuum sources, with flux densities of $\gtrsim 1~{\rm mJy}$.

\begin{figure}[hbt]
\begin{center}
\resizebox{3.25in}{!}{\includegraphics{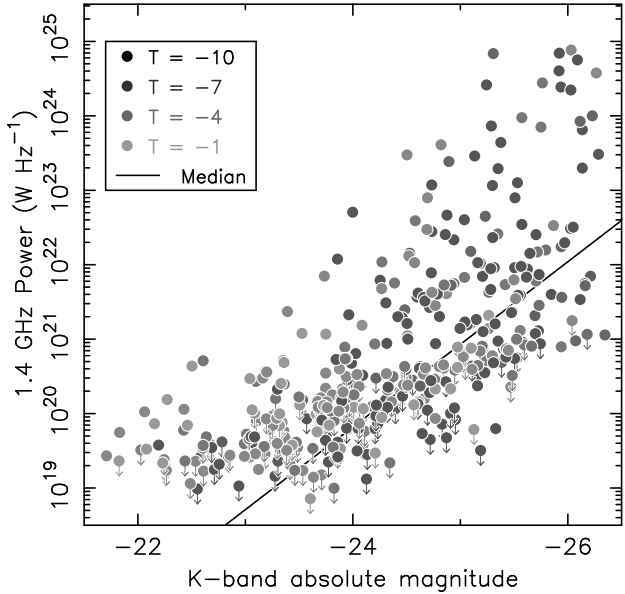}}
\end{center}
\caption{The $1.4~{\rm GHz}$ radio power of early-type galaxies as a function of $K$-band absolute
magnitude. The data are color coded by RC3 T type and for sources with measured flux densities less 
than $2\sigma$ above zero, we plot $2\sigma$ upper limits. At a fixed absolute magnitude, the distribution
of radio powers spans four orders of magnitude for early-type galaxies brighter than $M_K = -24$.
The median radio power of early-type galaxies (derived from fitting a log-normal probability density function to 
the data) is proportional to $K$-band luminosity to the power of  $2.78 \pm 0.16$.}
\vspace*{0.5cm}
\label{fig:lums}
\end{figure}

In Figure~\ref{fig:lums} we present the radio power of early-type galaxies as a function
of $K$-band absolute magnitude. As noted by others  \citep[e.g.,][]{fab89,sad89}, 
the radio powers of early-type galaxies are a strong function of absolute magnitude. 
Although our $K=9$ magnitude limit excludes some extremely powerful radio sources (e.g., Cygnus A),
the radio power of $M_K<-24$ galaxies at fixed $K$-band magnitude (or stellar mass) still spans 4 orders of magnitude. 
The most massive galaxies can have $1.4~{\rm GHz}$ powers as large as that of M~87 ($\simeq 7\times 10^{24}~{\rm W~Hz^{-1}}$) 
or less than the Milky Way ($\simeq 4\times 10^{21}~{\rm W~Hz^{-1}}$).

To model the distribution of radio powers, $P_{1.4}$, we have utilized a log-normal probability density function where
\begin{equation}
\rho (P_{1.4};\mu,\sigma) = \frac{1}{P_{1.4}\sigma\sqrt{2\pi}}{\rm exp}\left[-\frac{({\rm ln} P_{1.4} - \mu)^2}{2\sigma^2}\right]
\end{equation}
\begin{equation}
\mu= {\rm ln} \left[\alpha \left( \frac{L_K}{10^{11}L_\odot}\right)^\beta \right]
\end{equation}
\begin{equation}
\sigma = \gamma \mu
\end{equation}
where $L_K$ is the $K$-band luminosity and $\alpha$, $\beta$ and $\gamma$ are free parameters. 
This parameterization is empirical (i.e., lacks physical motivation) but, as discussed below, provides
a good description of the observed distribution of radio powers as a function of $L_K$.
If the radio emission from early-type galaxies was a function of $L_K$ and time only, then 
this parameterization would directly measure the radio duty cycle of early-type galaxies.
However, the radio emission from early-type galaxies is a function of environment 
\citep[host halo mass and location within a halo; e.g.,][]{bes07,wak08,man09}, so this 
parameterization constrains rather than measures the duty cycle. 
The integral of the probability density function provides an estimate of the fraction of galaxies with 
luminosity $L_K$ that have radio powers above (or below) a particular threshold, and this can be compared
with estimates from the prior literature \citep[e.g.,][]{bes05}. The radio luminosity function can be derived by 
convolving the probability density function with the $K$-band luminosity function of early-type galaxies.

We used our model of the distribution of radio powers and an empirical model of the NVSS flux density 
measurement errors to determine the likelihood of a particular galaxy (with known luminosity distance and $M_K$) 
having a particular measured flux density. Our model of the NVSS flux density measurement errors is the distribution of 
flux densities measured in each pixel of the NVSS, so we account for the non-Gaussian error 
distribution and source confusion. We then used the maximum likelihood method to determine the 
best-fit values and uncertainties for our model of the radio powers of early-type galaxies.
All galaxies were used when fitting the model, irrespective of their measured flux densities.
However, $M_K>-23$ galaxies provide limited constraints on the model parameters so our conclusions 
principally apply to $M_K<-23$ early-type galaxies. 
Our best-fit values for the log-normal distribution parameters $\alpha$, $\beta$ and $\gamma$ are 
$1.16 \pm 0.20 \times 10^{20}~{\rm W~Hz^{-1}}$, $2.78 \pm 0.17$ and $6.62\pm 0.29 \times 10^{-2}$ respectively.

\begin{figure}[hbt]
\begin{center}
\resizebox{3.25in}{!}{\includegraphics{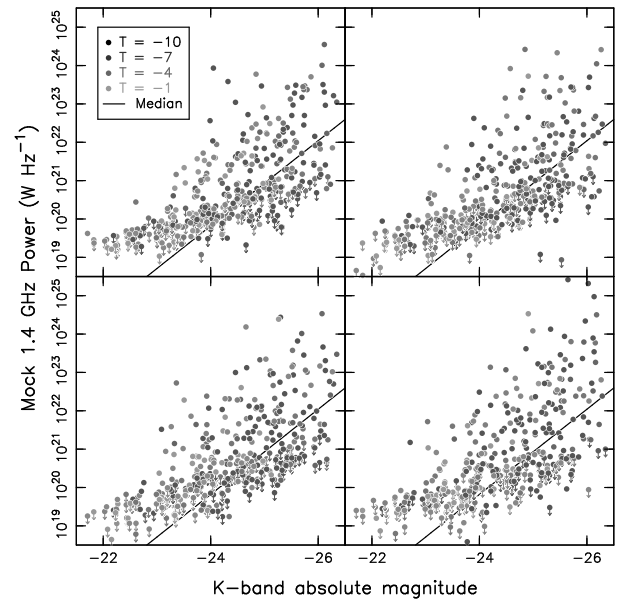}}
\end{center}
\caption{The radio powers of early-type galaxies as a function of $K$-band 
absolute magnitude for four mock catalogs. The distribution of radio powers approximates 
what is seen in Figure~\ref{fig:lums}. Apparent features are seen in 
the real and mock catalogs, but these are not significant departures from 
our model of the distribution of radio powers.}
\vspace*{0.5cm}
\label{fig:mock}
\end{figure}

To verify our model, we have used it to generate 500 realizations of the anticipated observed properties of our sample, 
and four of these ``mock catalogs''  are plotted in Figure~\ref{fig:mock}. 2D Kolmogorov-Smirnov (KS) tests \citep{pea83} run on 
the mocks and the original sample show that the our model is consistent 
with the observed distribution of radio powers. We also used the mocks to test the assumption that the 
measured minimum flux density is an underestimate of the true minimum flux density for $M_K<-25.5$ 
early-type galaxies (due to random errors broadening the measured flux density distribution).
For 96\% of the mock catalogs, the measured minimum $1.4~{\rm GHz}$ flux density is less 
than the true minimum $1.4~{\rm GHz}$ flux density for $M_K<-25.5$ early-type galaxies. It is thus 
is highly likely that all $M_K<-25.5$ early-type galaxies are radio continuum sources. 

The median $1.4~{\rm GHz}$ radio power of early-type galaxies is proportional to $K$-band luminosity to the power of $2.78 \pm 0.17$.
For comparison, \cite{fab89} and \cite{sad89} find $5~{\rm GHz}$ radio power is proportional to $B$-band luminosity to the 
power of $2.36\pm^{0.37}_{0.27}$ and $2.2\pm 0.3$ respectively. As these studies use $B$-band and $5~{\rm GHz}$ imagery, the modest 
differences between these results and our own may arise from the different wavelengths used. 
The fraction of early-type galaxies hosting a radio source 
more powerful than $10^{23}~{\rm W~Hz^{-1}}$ is roughly proportional to $L_K^{2.2}$, increasing from 
$0.2\%$ at $M_K=-23.5$ to $26\%$ at $M_K=-26$, in good agreement with \citet{mau07}. The observed trend 
also agrees well with the radio-loud fraction as a function of stellar mass derived  by \citet{bes05}, 
if $M_K=-26$ corresponds to a stellar mass of $\simeq 4\times 10^{11}~M_\odot$ \citep[e.g.,][]{bel01}. 
As the relationship between radio power and $K$-band absolute magnitude is extremely steep and the mass function of galaxies has an 
exponential cut-off at high masses, the most luminous radio sources will be hosted by galaxies spanning a 
relatively small range of $K$-band absolute magnitude. 

The distribution of radio powers at fixed $K$-band absolute magnitude is extremely broad, 
spanning approximately 4 orders of magnitude. A consequence of this is that stacking of radio sources
will converge slower towards the mean or median value than one would expect for a narrower Gaussian 
distribution of radio powers. A stack of 100 early-type galaxies will result in an estimate
of the mean radio power that is accurate to only $\simeq 20\%$, even if individual galaxies are detected 
with high signal-to-noise.

While it is beyond the scope of this letter to explore all the possibilities that could influence 
radio power, we can briefly explore some of the obvious options. At a given $K$-band absolute
magnitude, there is no clear correlation between radio power and morphological type in Figure~\ref{fig:lums}, 
and this conclusion is consistent with previous studies of early-type galaxies \citep[e.g.,][]{wro91}.
For $M_K<-25$ early-type galaxies, the average RC3 T types of galaxies that are above and below 
the median radio power differ by just $0.6$. More subtle correlations between morphology 
(including transitory features) and radio power may only be revealed by high resolution  
imaging of early-type galaxies \citep[e.g.,][]{der05}.

While others have explored the correlations between radio power and 
environment (e.g., dark matter halo mass) in detail, two galaxies in our sample illustrate 
that any correlation between environment and radio power is likely to have considerable 
scatter. Coma cluster galaxies NGC~4874 and NGC~4889 (the brightest cluster galaxy) both 
have $M_K\simeq -26.2$ and RC3 T Types of -4, but their $1.4~{\rm GHz}$ flux densities
are  $724~{\rm mJy}$ and $1~{\rm mJy}$ respectively. This large difference in radio flux density 
for two otherwise similar galaxies suggests that the radio powers of early-type galaxies may 
vary by roughly than 3 orders of magnitude over long periods of time. 

\section{Weak Radio Sources in Early-Type Galaxies}

\begin{figure}[hbt]
\begin{center}
\resizebox{3.25in}{!}{\includegraphics{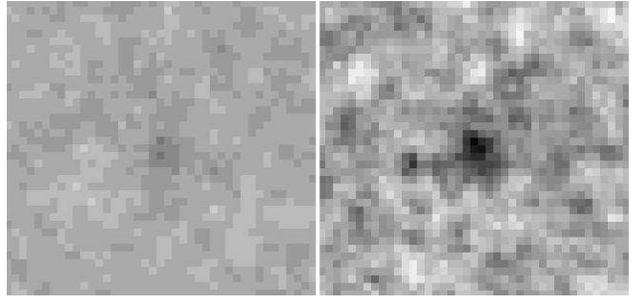}}
\end{center}
\caption{Median stacks of early-type galaxies with $M_K>-24$ (left) and  $M_K<-25.5$ (right) 
without counterparts in the NVSS catalogs. The images are $10^{\prime}$ on a side, the pixel scale
is $15^{\prime\prime}$, white corresponds to $-0.5~{\rm mJy}$ and black corresponds to $1~{\rm mJy}$.  
While little radio emission is detected from the low mass galaxies, the stack of high mass galaxies 
has a flux density of $0.9~{\rm mJy}$.}
\label{fig:stack}
\end{figure}

We have previously concluded that all $M_K<-25.5$ early-type galaxies are probably
radio sources, as their measured flux densities are greater than zero. 
To verify the presence of weak radio sources in massive galaxies, we have stacked NVSS images of the 
16 $M_K<-25.5$ early-type galaxies without counterparts in the NVSS catalogs. For comparison, we have 
also stacked images of the 119 $M_K>-24$ early-type galaxies without counterparts in the NVSS catalogs. 
As we show in Figure~\ref{fig:stack}, a $0.9~{\rm mJy}$ source is present in the stack of massive
galaxies while radio emission is only marginally detected in the stack of lower mass galaxies. This 
is not unexpected, as the median radio power of early-type galaxies is an extremely strong function
$K$-band absolute magnitude. 
The radio continuum emission from $M_K<-25.5$ early-type galaxies indicates that they are rarely (perhaps never)  
truly passive, as the vast majority harbor an AGN or have recently ($\lesssim 100~{\rm Myr}$) undergone star formation.

The most powerful radio sources in our sample are clearly AGNs, but the weakest radio sources could result
from star formation. Star formation is known to produce $10^{21}~{\rm W~Hz^{-1}}$ of radio emission 
in some early-type galaxies \citep{wro88}, which would result from  just $\sim 1~M_\odot/{\rm yr}$ of star 
formation \citep{bel03}. However, there are weak arguments for AGNs being responsible for the broad distribution
of observed radio powers for $M_K<-25.5$ early-type galaxies. The radio emission in the stacked images is compact, 
as one may expect from low luminosity AGNs, although the NVSS beam is $45^{\prime\prime}$ wide. \citet{sle94} 
found parsec-sized radio cores in 70\% of early-type galaxies, even when the total radio power was only $10^{22}~{\rm W~Hz^{-1}}$.
If radio emission is being produced by multiple mechanisms, one may expect an obvious superposition of multiple distributions. 
This should be the case for X-ray emission from early-type galaxies \citep[e.g.,][]{fab89}, but we do not see this in 
the radio in Figure~\ref{fig:lums}.  However, these are not definitive arguments for 
the presence of radio AGNs in all massive galaxies and further observations are required. 
Inconsistencies between star formation rates derived from observations at radio and other wavelengths (e.g., mid-IR)
may identify the weakest AGNs in early-type galaxies. Alternatively, AGNs may be identified using 
source sizes and brightness temperatures determined with high frequency and high resolution radio imaging.

\section{Summary}

We have measured the $1.4~{\rm GHz}$ radio continuum emission from $K<9$ early-type galaxies, 
utilizing archival imagery from the NVSS, Green Bank 300-ft Telescope and 64-m Parkes Radio Telescope. 
The distribution of radio powers at fixed absolute magnitude spans 4 orders of magnitude 
for $M_K<-24$ early-type galaxies, and the median radio power is proportional to $K$-band luminosity 
to the power $2.78\pm 0.16$.
Our analysis is not restricted to galaxies with high signal-to-noise radio detections, 
and the distribution of noisy flux density measurements provides important constraints 
on the radio emission from early-type galaxies. Relatively few low mass early-type
galaxies are detected with high signal-to-noise, and the flux density measurements for the remaining
objects are scattered around zero. In contrast, most $M_K<-25.5$ early-type galaxies have robust 
detections, and the remaining galaxies consistently have flux density measurements greater than zero. 
As random measurement errors will broaden the measured distribution of flux densities, it
is highly likely that all massive galaxies are radio continuum sources. If this is the case, 
all $M_K<25.5$ early-type galaxies harbor an active galactic nucleus or have recently undergone star formation.

\begin{acknowledgments}
The National Radio Astronomy Observatory is a facility of the National Science Foundation 
operated under cooperative agreement by Associated Universities, Inc. The Parkes telescope is 
part of the Australia Telescope National Facility which is funded by the Commonwealth of 
Australia for operation as a National Facility managed by CSIRO. We thank the HIPASS 
collaboration and Mark Calabretta for making their radio continuum maps available to us prior to publication. 
This publication makes use of data products from the Two Micron All Sky Survey, which is a 
joint project of the University of Massachusetts and the Infrared Processing and Analysis 
Center/California Institute of Technology, funded by the National Aeronautics and Space 
Administration and the National Science Foundation. The development of this letter was
assisted by discussions with many astronomers, including Geoffrey Bower, Scott Croom, 
Steve Croft, Arjun Dey and Elaine Sadler.
\end{acknowledgments}

\bibliographystyle{astroads}
\bibliography{ms.bib}

\begin{thebibliography}{42}
\expandafter\ifx\csname natexlab\endcsname\relax\def\natexlab#1{#1}\fi
\expandafter\ifx\csname href\endcsname\relax
  \def\href#1#2{}\fi
\expandafter\ifx\csname urllinklabel\endcsname\relax
  \def\urllinklabel{[LINK]}\fi
\expandafter\ifx\csname adsurllinklabel\endcsname\relax
  \def\adsurllinklabel{[ADS]}\fi

\bibitem[{{Baldi} \& {Capetti}(2009)}]{bal09}
{Baldi}, R.~D. \& {Capetti}, A. 2009, \aap, 508, 603


\bibitem[{{Barnes} {et~al.}(2001){Barnes}, {Staveley-Smith}, {de Blok},
  {Oosterloo}, {Stewart}, {Wright}, {Banks}, {Bhathal}, {Boyce}, {Calabretta},
  {Disney}, {Drinkwater}, {Ekers}, {Freeman}, {Gibson}, {Green}, {Haynes}, {te
  Lintel Hekkert}, {Henning}, {Jerjen}, {Juraszek}, {Kesteven}, {Kilborn},
  {Knezek}, {Koribalski}, {Kraan-Korteweg}, {Malin}, {Marquarding}, {Minchin},
  {Mould}, {Price}, {Putman}, {Ryder}, {Sadler}, {Schr{\"o}der}, {Stootman},
  {Webster}, {Wilson}, \& {Ye}}]{bar01}
{Barnes}, D.~G., {Staveley-Smith}, L., {de Blok}, W.~J.~G., {Oosterloo}, T.,
  {Stewart}, I.~M., {Wright}, A.~E., {Banks}, G.~D., {Bhathal}, R., {Boyce},
  P.~J., {Calabretta}, M.~R., {Disney}, M.~J., {Drinkwater}, M.~J., {Ekers},
  R.~D., {Freeman}, K.~C., {Gibson}, B.~K., {Green}, A.~J., {Haynes}, R.~F.,
  {te Lintel Hekkert}, P., {Henning}, P.~A., {Jerjen}, H., {Juraszek}, S.,
  {Kesteven}, M.~J., {Kilborn}, V.~A., {Knezek}, P.~M., {Koribalski}, B.,
  {Kraan-Korteweg}, R.~C., {Malin}, D.~F., {Marquarding}, M., {Minchin}, R.~F.,
  {Mould}, J.~R., {Price}, R.~M., {Putman}, M.~E., {Ryder}, S.~D., {Sadler},
  E.~M., {Schr{\"o}der}, A., {Stootman}, F., {Webster}, R.~L., {Wilson}, W.~E.,
  \& {Ye}, T. 2001, \mnras, 322, 486


\bibitem[{{Bell}(2003)}]{bel03}
{Bell}, E.~F. 2003, \apj, 586, 794


\bibitem[{{Bell} \& {de Jong}(2001)}]{bel01}
{Bell}, E.~F. \& {de Jong}, R.~S. 2001, \apj, 550, 212


\bibitem[{{Best} {et~al.}(2005){Best}, {Kauffmann}, {Heckman}, {Brinchmann},
  {Charlot}, {Ivezi{\'c}}, \& {White}}]{bes05}
{Best}, P.~N., {Kauffmann}, G., {Heckman}, T.~M., {Brinchmann}, J., {Charlot},
  S., {Ivezi{\'c}}, {\v Z}., \& {White}, S.~D.~M. 2005, \mnras, 362, 25


\bibitem[{{Best} {et~al.}(2007){Best}, {von der Linden}, {Kauffmann},
  {Heckman}, \& {Kaiser}}]{bes07}
{Best}, P.~N., {von der Linden}, A., {Kauffmann}, G., {Heckman}, T.~M., \&
  {Kaiser}, C.~R. 2007, \mnras, 379, 894


\bibitem[{{Birnboim} \& {Dekel}(2003)}]{bir03}
{Birnboim}, Y. \& {Dekel}, A. 2003, \mnras, 345, 349


\bibitem[{{Brown} {et~al.}(2008){Brown}, {Zheng}, {White}, {Dey}, {Jannuzi},
  {Benson}, {Brand}, {Brodwin}, \& {Croton}}]{bro08}
{Brown}, M.~J.~I., {Zheng}, Z., {White}, M., {Dey}, A., {Jannuzi}, B.~T.,
  {Benson}, A.~J., {Brand}, K., {Brodwin}, M., \& {Croton}, D.~J. 2008, \apj,
  682, 937


\bibitem[{{Canizares} {et~al.}(1987){Canizares}, {Fabbiano}, \&
  {Trinchieri}}]{can87}
{Canizares}, C.~R., {Fabbiano}, G., \& {Trinchieri}, G. 1987, \apj, 312, 503


\bibitem[{{Condon}(1992)}]{con92}
{Condon}, J.~J. 1992, \araa, 30, 575


\bibitem[{{Condon} \& {Broderick}(1985)}]{con85}
{Condon}, J.~J. \& {Broderick}, J.~J. 1985, \aj, 90, 2540


\bibitem[{{Condon} \& {Broderick}(1986)}]{con86}
---. 1986, \aj, 91, 1051


\bibitem[{{Condon} {et~al.}(1998){Condon}, {Cotton}, {Greisen}, {Yin},
  {Perley}, {Taylor}, \& {Broderick}}]{con98}
{Condon}, J.~J., {Cotton}, W.~D., {Greisen}, E.~W., {Yin}, Q.~F., {Perley},
  R.~A., {Taylor}, G.~B., \& {Broderick}, J.~J. 1998, \aj, 115, 1693


\bibitem[{{Croton} {et~al.}(2006){Croton}, {Springel}, {White}, {De Lucia},
  {Frenk}, {Gao}, {Jenkins}, {Kauffmann}, {Navarro}, \& {Yoshida}}]{cro06}
{Croton}, D.~J., {Springel}, V., {White}, S.~D.~M., {De Lucia}, G., {Frenk},
  C.~S., {Gao}, L., {Jenkins}, A., {Kauffmann}, G., {Navarro}, J.~F., \&
  {Yoshida}, N. 2006, \mnras, 365, 11


\bibitem[{{de Ruiter} {et~al.}(2005){de Ruiter}, {Parma}, {Capetti}, {Fanti},
  {Morganti}, \& {Santantonio}}]{der05}
{de Ruiter}, H.~R., {Parma}, P., {Capetti}, A., {Fanti}, R., {Morganti}, R., \&
  {Santantonio}, L. 2005, \aap, 439, 487


\bibitem[{{de Vaucouleurs} {et~al.}(1991){de Vaucouleurs}, {de Vaucouleurs},
  {Corwin}, {Buta}, {Paturel}, \& {Fouque}}]{rc3}
{de Vaucouleurs}, G., {de Vaucouleurs}, A., {Corwin}, Jr., H.~G., {Buta},
  R.~J., {Paturel}, G., \& {Fouque}, P. 1991, {Third Reference Catalogue of
  Bright Galaxies}


\bibitem[{{Fabbiano} {et~al.}(1989){Fabbiano}, {Gioia}, \&
  {Trinchieri}}]{fab89}
{Fabbiano}, G., {Gioia}, I.~M., \& {Trinchieri}, G. 1989, \apj, 347, 127


\bibitem[{{Fabian} {et~al.}(2003){Fabian}, {Sanders}, {Allen}, {Crawford},
  {Iwasawa}, {Johnstone}, {Schmidt}, \& {Taylor}}]{fab03a}
{Fabian}, A.~C., {Sanders}, J.~S., {Allen}, S.~W., {Crawford}, C.~S.,
  {Iwasawa}, K., {Johnstone}, R.~M., {Schmidt}, R.~W., \& {Taylor}, G.~B. 2003,
  \mnras, 344, L43


\bibitem[{{Falco} {et~al.}(1999){Falco}, {Kurtz}, {Geller}, {Huchra}, {Peters},
  {Berlind}, {Mink}, {Tokarz}, \& {Elwell}}]{fal99}
{Falco}, E.~E., {Kurtz}, M.~J., {Geller}, M.~J., {Huchra}, J.~P., {Peters}, J.,
  {Berlind}, P., {Mink}, D.~J., {Tokarz}, S.~P., \& {Elwell}, B. 1999, \pasp,
  111, 438


\bibitem[{{Fanaroff} \& {Riley}(1974)}]{far74}
{Fanaroff}, B.~L. \& {Riley}, J.~M. 1974, \mnras, 167, 31P


\bibitem[{{Hopkins} {et~al.}(2006){Hopkins}, {Hernquist}, {Cox}, {Di Matteo},
  {Robertson}, \& {Springel}}]{hop06}
{Hopkins}, P.~F., {Hernquist}, L., {Cox}, T.~J., {Di Matteo}, T., {Robertson},
  B., \& {Springel}, V. 2006, \apjs, 163, 1


\bibitem[{{Huchra} {et~al.}(1999){Huchra}, {Vogeley}, \& {Geller}}]{huc99}
{Huchra}, J.~P., {Vogeley}, M.~S., \& {Geller}, M.~J. 1999, \apjs, 121, 287


\bibitem[{{Jarrett} {et~al.}(2000){Jarrett}, {Chester}, {Cutri}, {Schneider},
  {Skrutskie}, \& {Huchra}}]{jar00}
{Jarrett}, T.~H., {Chester}, T., {Cutri}, R., {Schneider}, S., {Skrutskie}, M.,
  \& {Huchra}, J.~P. 2000, \aj, 119, 2498


\bibitem[{{Jones} {et~al.}(2009){Jones}, {Read}, {Saunders}, {Colless},
  {Jarrett}, {Parker}, {Fairall}, {Mauch}, {Sadler}, {Watson}, {Burton},
  {Campbell}, {Cass}, {Croom}, {Dawe}, {Fiegert}, {Frankcombe}, {Hartley},
  {Huchra}, {James}, {Kirby}, {Lahav}, {Lucey}, {Mamon}, {Moore}, {Peterson},
  {Prior}, {Proust}, {Russell}, {Safouris}, {Wakamatsu}, {Westra}, \&
  {Williams}}]{jon09}
{Jones}, D.~H., {Read}, M.~A., {Saunders}, W., {Colless}, M., {Jarrett}, T.,
  {Parker}, Q.~A., {Fairall}, A.~P., {Mauch}, T., {Sadler}, E.~M., {Watson},
  F.~G., {Burton}, D., {Campbell}, L.~A., {Cass}, P., {Croom}, S.~M., {Dawe},
  J., {Fiegert}, K., {Frankcombe}, L., {Hartley}, M., {Huchra}, J., {James},
  D., {Kirby}, E., {Lahav}, O., {Lucey}, J., {Mamon}, G.~A., {Moore}, L.,
  {Peterson}, B.~A., {Prior}, S., {Proust}, D., {Russell}, K., {Safouris}, V.,
  {Wakamatsu}, K., {Westra}, E., \& {Williams}, M. 2009, \mnras, 399, 683


\bibitem[{{Mandelbaum} {et~al.}(2009){Mandelbaum}, {Li}, {Kauffmann}, \&
  {White}}]{man09}
{Mandelbaum}, R., {Li}, C., {Kauffmann}, G., \& {White}, S.~D.~M. 2009, \mnras,
  393, 377


\bibitem[{{Mauch} \& {Sadler}(2007)}]{mau07}
{Mauch}, T. \& {Sadler}, E.~M. 2007, \mnras, 375, 931


\bibitem[{{McNamara} {et~al.}(2000){McNamara}, {Wise}, {Nulsen}, {David},
  {Sarazin}, {Bautz}, {Markevitch}, {Vikhlinin}, {Forman}, {Jones}, \&
  {Harris}}]{mcn00}
{McNamara}, B.~R., {Wise}, M., {Nulsen}, P.~E.~J., {David}, L.~P., {Sarazin},
  C.~L., {Bautz}, M., {Markevitch}, M., {Vikhlinin}, A., {Forman}, W.~R.,
  {Jones}, C., \& {Harris}, D.~E. 2000, \apjl, 534, L135


\bibitem[{{Peacock}(1983)}]{pea83}
{Peacock}, J.~A. 1983, \mnras, 202, 615


\bibitem[{{Sadler} {et~al.}(1989){Sadler}, {Jenkins}, \& {Kotanyi}}]{sad89}
{Sadler}, E.~M., {Jenkins}, C.~R., \& {Kotanyi}, C.~G. 1989, \mnras, 240, 591


\bibitem[{{Schlegel} {et~al.}(1998){Schlegel}, {Finkbeiner}, \&
  {Davis}}]{sch98}
{Schlegel}, D.~J., {Finkbeiner}, D.~P., \& {Davis}, M. 1998, \apj, 500, 525


\bibitem[{{Shabala} {et~al.}(2008){Shabala}, {Ash}, {Alexander}, \&
  {Riley}}]{sha08}
{Shabala}, S.~S., {Ash}, S., {Alexander}, P., \& {Riley}, J.~M. 2008, \mnras,
  388, 625


\bibitem[{{Slee} {et~al.}(1994){Slee}, {Sadler}, {Reynolds}, \&
  {Ekers}}]{sle94}
{Slee}, O.~B., {Sadler}, E.~M., {Reynolds}, J.~E., \& {Ekers}, R.~D. 1994,
  \mnras, 269, 928


\bibitem[{{Spergel} {et~al.}(2007){Spergel}, {Bean}, {Dor{\'e}}, {Nolta},
  {Bennett}, {Dunkley}, {Hinshaw}, {Jarosik}, {Komatsu}, {Page}, {Peiris},
  {Verde}, {Halpern}, {Hill}, {Kogut}, {Limon}, {Meyer}, {Odegard}, {Tucker},
  {Weiland}, {Wollack}, \& {Wright}}]{spe07}
{Spergel}, D.~N., {Bean}, R., {Dor{\'e}}, O., {Nolta}, M.~R., {Bennett}, C.~L.,
  {Dunkley}, J., {Hinshaw}, G., {Jarosik}, N., {Komatsu}, E., {Page}, L.,
  {Peiris}, H.~V., {Verde}, L., {Halpern}, M., {Hill}, R.~S., {Kogut}, A.,
  {Limon}, M., {Meyer}, S.~S., {Odegard}, N., {Tucker}, G.~S., {Weiland},
  J.~L., {Wollack}, E., \& {Wright}, E.~L. 2007, \apjs, 170, 377


\bibitem[{{Tabor} \& {Binney}(1993)}]{tab93}
{Tabor}, G. \& {Binney}, J. 1993, \mnras, 263, 323


\bibitem[{{Tinsley}(1968)}]{tin68}
{Tinsley}, B.~M. 1968, \apj, 151, 547


\bibitem[{{Trager} {et~al.}(2000){Trager}, {Faber}, {Worthey}, \&
  {Gonz{\'a}lez}}]{tra00}
{Trager}, S.~C., {Faber}, S.~M., {Worthey}, G., \& {Gonz{\'a}lez}, J.~J. 2000,
  \aj, 120, 165


\bibitem[{{Tully} {et~al.}(2009){Tully}, {Rizzi}, {Shaya}, {Courtois},
  {Makarov}, \& {Jacobs}}]{tul09}
{Tully}, R.~B., {Rizzi}, L., {Shaya}, E.~J., {Courtois}, H.~M., {Makarov},
  D.~I., \& {Jacobs}, B.~A. 2009, \aj, 138, 323


\bibitem[{{Wake} {et~al.}(2008){Wake}, {Croom}, {Sadler}, \&
  {Johnston}}]{wak08}
{Wake}, D.~A., {Croom}, S.~M., {Sadler}, E.~M., \& {Johnston}, H.~M. 2008,
  \mnras, 391, 1674


\bibitem[{{Wegner} {et~al.}(2003){Wegner}, {Bernardi}, {Willmer}, {da Costa},
  {Alonso}, {Pellegrini}, {Maia}, {Chaves}, \& {Rit{\'e}}}]{weg03}
{Wegner}, G., {Bernardi}, M., {Willmer}, C.~N.~A., {da Costa}, L.~N., {Alonso},
  M.~V., {Pellegrini}, P.~S., {Maia}, M.~A.~G., {Chaves}, O.~L., \& {Rit{\'e}},
  C. 2003, \aj, 126, 2268


\bibitem[{{White} \& {Rees}(1978)}]{whi78}
{White}, S.~D.~M. \& {Rees}, M.~J. 1978, \mnras, 183, 341


\bibitem[{{Wrobel} \& {Heeschen}(1988)}]{wro88}
{Wrobel}, J.~M. \& {Heeschen}, D.~S. 1988, \apj, 335, 677


\bibitem[{{Wrobel} \& {Heeschen}(1991)}]{wro91}
---. 1991, \aj, 101, 148


\end{thebibliography}

\begin{deluxetable}{cccccccc}
\tablecolumns{8}
\%
\tablecaption{The $K<9$ Early-Type Galaxy Sample (complete table is available online)\label{table:summary}}
\tablehead{
  \colhead{Name}                   &
  \colhead{J2000 Coordinates}      &
  \colhead{2MASS $m_K$}            &
   \colhead{RC3 T Type}             &
  \colhead{$S_{1.4}$}               &
  \colhead{$d_L$}                  &
  \colhead{$M_K$}                  &
  \colhead{$P_{1.4}$\tablenotemark{a}} \\
  \colhead{}                &
  \colhead{}                &
  \colhead{(Dust corrected)}                &
  \colhead{}                &
  \colhead{(mJy)}           &
  \colhead{(Mpc)}           &
  \colhead{}                &
  \colhead{$(\rm W~Hz^{-1})$}}
\startdata               
         NGC    16    &  00:09:04.300 +27:43:46.03  &  8.76 & -3.0 & $  -1.0 \pm  0.5 $               &    42 & -24.34 & -                   \\  
         NGC    50    &  00:14:44.555 -07:20:42.38  &  8.66 & -3.0 & $    22 \pm    1 $               &    75 & -25.72 & $    1.5 \times 10^{22}  $ \\  
         NGC    57    &  00:15:30.873 +17:19:42.22  &  8.65 & -5.0 & $   0.9 \pm  0.5 $               &    74 & -25.70 & $ <  1.2 \times 10^{21}  $ \\  
         NGC    80    &  00:21:10.865 +22:21:26.11  &  8.90 & -2.5 & $   0.9 \pm  0.5 $               &    78 & -25.56 & $ <  1.3 \times 10^{21}  $ \\  
         NGC   128    &  00:29:15.070 +02:51:50.57  &  8.51 & -2.0 & $   1.5 \pm  0.5 $               &    58 & -25.30 & $    6.0 \times 10^{20}  $ \\  
         NGC   205    &  00:40:22.075 +41:41:07.08  &  5.56 & -5.0 & $   0.1 \pm  0.5 $               &   0.8 & -18.84 & $ <  6.9 \times 10^{16}  $ \\  
         NGC   221    &  00:42:41.825 +40:51:54.61  &  5.05 & -6.0 & $   0.7 \pm  0.5 $               &   0.8 & -19.50 & $ <  1.3 \times 10^{17}  $ \\  
         NGC   315    &  00:57:48.916 +30:21:08.33  &  7.93 & -4.0 & $   1.8 \pm  0.1 \times 10^{3} $ &    68 & -26.23 & $    1.0 \times 10^{24}  $ \\  
         NGC   383    &  01:07:24.939 +32:24:45.20  &  8.46 & -3.0 & $   4.8 \pm  0.2 \times 10^{3} $ &    70 & -25.76 & $    2.8 \times 10^{24}  $ \\  
         NGC   410    &  01:10:58.872 +33:09:07.30  &  8.36 & -4.0 & $   5.8 \pm  0.5 $               &    73 & -25.94 & $    3.6 \times 10^{21}  $ \\  
         NGC   439    &  01:13:47.251 -31:44:50.09  &  8.68 & -3.3 & $    21 \pm    1 $               &    79 & -25.79 & $    1.6 \times 10^{22}  $ \\  
         NGC   474    &  01:20:06.696 +03:24:54.97  &  8.54 & -2.0 & $   0.3 \pm  0.5 $               &    32 & -23.97 & $ <  1.5 \times 10^{20}  $ \\  
         NGC   499    &  01:23:11.459 +33:27:36.30  &  8.71 & -2.5 & $   0.7 \pm  0.5 $               &    60 & -25.18 & $ <  6.9 \times 10^{20}  $ \\  
         NGC   507    &  01:23:39.950 +33:15:22.22  &  8.28 & -2.0 & $    62 \pm    2 $               &    67 & -25.87 & $    3.4 \times 10^{22}  $ \\  
         NGC   524    &  01:24:47.707 +09:32:19.65  &  7.13 & -1.0 & $   3.1 \pm  0.4 $               &    24 & -24.77 & $    2.1 \times 10^{20}  $ \\  
         NGC   533    &  01:25:31.432 +01:45:33.57  &  8.43 & -5.0 & $    29 \pm    1 $               &    76 & -25.97 & $    2.0 \times 10^{22}  $ \\  
         NGC   547    &  01:26:00.577 -01:20:42.43  &  8.48 & -5.0 & $   5.9 \pm  0.5 \times 10^{3} $ &    76 & -25.92 & $    4.0 \times 10^{24}  $ \\  
         NGC   584    &  01:31:20.755 -06:52:05.02  &  7.29 & -5.0 & $   0.6 \pm  0.5 $               &    20 & -24.23 & $ <  7.3 \times 10^{19}  $ \\  
         NGC   596    &  01:32:51.906 -07:01:53.54  &  7.96 & -4.0 & $   0.1 \pm  0.5 $               &    22 & -23.73 & $ <  5.7 \times 10^{19}  $ \\  
         NGC   636    &  01:39:06.529 -07:30:45.37  &  8.43 & -5.0 & $  -0.3 \pm  0.5 $               &    30 & -23.94 & $ <  6.4 \times 10^{19}  $ \\  
\hline
\enddata
\tablenotetext{a}{When the flux density is within $2\sigma$ of zero, a $2\sigma$ upper limit for the radio power is provided.}
\end{deluxetable}

\end{document}